\documentclass[conference]{IEEEtran}
\IEEEoverridecommandlockouts
\usepackage{cite}
\usepackage{amsmath,amssymb,amsfonts}
\usepackage{algorithmic}
\usepackage{graphicx}
\usepackage{textcomp}
\usepackage{xcolor}
\usepackage{tabularx,booktabs}

\def\BibTeX{{\rm B\kern-.05em{\sc i\kern-.025em b}\kern-.08em
    T\kern-.1667em\lower.7ex\hbox{E}\kern-.125emX}}
\begin{document}

\title{Entropy-Adaptive Multi-Map Chaotic Modulation for Physical-Layer Security

}

\author{\IEEEauthorblockN{Dhrumil Bhatt}
\IEEEauthorblockA{
\textit{Manipal Institute of Technology}\\
\textit{Manipal Academy of Higher Education}\\
Manipal, India \\
dhrumil.bhatt@gmail.com}
\and
\IEEEauthorblockN{Aarish Patel}
\IEEEauthorblockA{
\textit{Manipal Institute of Technology}\\
\textit{Manipal Academy of Higher Education}\\
Manipal, India \\
aarishpatel30@gmail.com}}

\maketitle
\begin{abstract}
This paper presents a Multi-Map Dynamic-Entropy Intrusion-Aware Chaotic Modulation (MU-DE-IAEACM-MM) framework for adaptive physical-layer security in multi-user wireless systems. Unlike conventional chaos-based schemes that rely on static parameter secrecy, the proposed architecture treats entropy as a regulated security variable and dynamically updates chaotic control parameters using an entropy-driven adaptation law. Heterogeneous chaotic maps, including Logistic, Tent, Chebyshev, and Sine generators, are distributed across users to enlarge entropy dimensionality and reduce cross-user statistical dependence. A correlation-based intrusion metric is incorporated to detect improved adversarial reconstruction coherence and trigger controlled entropy escalation. Stability analysis establishes bounded convergence conditions for the adaptive update process. Monte Carlo simulations under AWGN, fading, impulsive, colored, and narrowband interference demonstrate a persistent BER gap between legitimate and mismatched receivers and measurable secrecy capacity gains over representative fixed-parameter chaotic modulation schemes. The framework maintains positive secrecy rates in dense deployments with up to 64 legitimate users and 30 passive eavesdroppers. The results indicate that entropy-regulated multi-map chaotic modulation provides a scalable and synchronisation-stable approach for adaptive physical-layer security in next-generation wireless and IoT networks.
\end{abstract}


\section{Introduction and Related Works}

The rapid expansion of dense wireless networks and large-scale Internet of Things (IoT) deployments has increased interest in lightweight physical-layer security mechanisms. In systems involving resource-constrained devices, real-time communication requirements, and large numbers of users, conventional upper-layer cryptographic techniques may introduce additional computational and latency overhead~\cite{yang2015safeguarding,lin2020lightweight,zhang2023survey}. As a result, waveform-level protection strategies have received renewed attention.

Chaos-based modulation embeds randomness directly into the transmitted waveform using nonlinear dynamical systems~\cite{feki2020chaos,alvarez2019review}. In these schemes, information symbols are mapped onto trajectories generated by chaotic maps such as Logistic, Tent, Chebyshev, or Sine functions. The resulting signals exhibit broadband spectral characteristics, strong sensitivity to initial conditions, and elevated entropy, which complicate reconstruction when precise system parameters are 
unknown~\cite{feki2020chaos,alvarez2019review}. Such properties make chaotic modulation attractive in short-range RF, ultra-wideband (UWB), and low-power IoT systems where hardware simplicity and waveform-level obfuscation are desirable.

However, most existing chaos-based implementations rely on fixed chaotic parameters and assume quasi-static operating conditions. Security is typically derived from parameter mismatch between legitimate transceivers and potential eavesdroppers. Under this common model, an adversary passively observes the transmitted waveform and attempts synchronization or parameter estimation using incomplete system knowledge. Although exact reconstruction remains difficult, statistical inference methods may gradually improve predictability if chaotic parameters remain unchanged over long transmissions.

Recent work has primarily focused on improving performance metrics such as spectral efficiency or noise robustness. PT-CIM-DCSK improves data rate and multipath robustness through parallel 
code-indexed chaotic transmission while retaining static chaotic 
parameters~\cite{b8}. NR-ODBR-DCSK improves BER performance and doubles the information transmission 
rate under AWGN using Walsh orthogonalization~\cite{b9}. RIS-assisted FM-DCSK adapts to channel conditions using reconfigurable intelligent surfaces but does not incorporate security-driven parameter evolution~\cite{huang2023rischaos}. Hardware demonstrations using software-defined radio platforms and UWB chaotic pulses confirm practical feasibility yet operate with fixed chaotic configurations~\cite{b13,b14}.
Other contributions explore non-coherent FM chaos-based modulation~\cite{huang2023rischaos},
chaos-intensity tuning for BER optimisation~\cite{rashid2022chaosnet}, and composite radar-communication waveforms~\cite{b16}. While these approaches improve reliability or spectral utilization, they do not explicitly regulate chaotic entropy in response to inference-based threats.

Two limitations remain evident. First, adversary models are often implicit and limited to static parameter mismatch assumptions. Second, multi-user systems commonly employ a single chaotic map family across users, which restricts entropy dimensionality and may introduce residual cross-correlation in dense deployments.

This work addresses the following problem: \textit{how to regulate chaotic signal entropy and parameter evolution in a bounded and stable manner to reduce waveform predictability under passive statistical inference, while preserving synchronization and supporting scalable multi-user operation}. The adversary considered in this study is a passive observer with imperfect parameter knowledge attempting synchronization or parameter estimation. Active jamming, injection, or signal manipulation attacks are outside the present scope.

The proposed Multi-Map Dynamic-Entropy Intrusion-Aware Chaotic Modulation framework differs from existing approaches in three key aspects. First, heterogeneous chaotic maps are distributed across users to increase entropy dimensionality and reduce inter-user statistical correlation. Second, chaotic control parameters evolve dynamically within provable stability bounds rather than remaining fixed. Third, entropy is treated as a regulated security variable and adjusted to limit long-term predictability under inference-based threat models, instead of being tuned solely for channel performance.

By moving from static parameter secrecy toward bounded entropy evolution, the proposed approach aims to strengthen chaotic modulation against passive statistical reconstruction while maintaining practical implementation feasibility in multi-user wireless and IoT systems.
\section{Methodology}

The proposed Multi-Map Dynamic-Entropy Intrusion-Aware Chaotic Modulation (MU-DE-IAEACM-MM) framework establishes a bounded, self-regulating chaotic communication architecture in which entropy is treated as an explicit security control variable. The framework integrates heterogeneous chaotic map generation, entropy-driven parameter adaptation, and correlation-based intrusion awareness within a passive-adversary model. Unlike conventional chaos-based systems that rely on static parameter secrecy, the proposed design continuously regulates chaotic complexity to suppress long-term statistical inference while preserving synchronisation stability.

Consider a multi-user wireless communication system comprising $U$ legitimate transmitter–receiver pairs and $E$ passive eavesdroppers. Each legitimate user $u \in \{1,\ldots,U\}$ employs a discrete-time chaotic generator $f_u(x;r_u)$ with control parameter $r_u$ and internal state $x_{u,n} \in (0,1)$ at time index $n$.

The state evolution follows

\begin{equation}
x_{u,n+1} = f_u(x_{u,n}; r_u),
\label{eq:state}
\end{equation}

where $f_u(\cdot)$ is selected from a family of nonlinear maps operating in a chaotic regime. The information symbol $m_{u,n} \in \{-1,+1\}$ modulates the chaotic trajectory through

\begin{equation}
s_{u,n} = m_{u,n} \phi_u(x_{u,n}),
\label{eq:waveform}
\end{equation}

where $\phi_u(\cdot)$ is a normalized shaping function ensuring
\[
\mathbb{E}[\phi_u^2(x)] = 1,
\]
thereby maintaining constant average symbol energy.

The transmitted composite signal is

\begin{equation}
s_n = \sum_{u=1}^{U} s_{u,n}.
\label{eq:composite}
\end{equation}

The chaotic maps are selected from

\begin{subequations}
\begin{align}
f(x;r) &= r x(1-x), \\
f(x;r) &= r \left(1 - 2|x-\tfrac{1}{2}|\right), \\
f(x;r) &= \cos(r \arccos(x)), \\
f(x;r) &= r \sin(\pi x),
\end{align}
\end{subequations}

corresponding respectively to Logistic, Tent, Chebyshev, and Sine maps. Each map is operated within parameter intervals known to produce positive Lyapunov exponents, ensuring sensitive dependence on initial conditions and parameter mismatch.

Heterogeneous map allocation increases entropy and dimensionality, and reduces inter-user statistical correlation, thereby mitigating parameter estimation via cross-user inference.

The chaotic parameter $r_u$ directly influences trajectory complexity and invariant distribution. To regulate unpredictability, the empirical Shannon entropy of the chaotic state over a sliding window is estimated as

\begin{equation}
H_u = -\sum_{k=1}^{K} p_k \log_2 p_k,
\label{eq:entropy}
\end{equation}

where $p_k$ denotes the probability of the state occupying the $k$th quantisation bin. Entropy serves as a proxy for waveform unpredictability and statistical flatness.

The adaptive update rule is

\begin{equation}
r_{u,n+1} = r_{u,n} + \alpha x_{u,n} + \mu (H_t(u) - H_u),
\label{eq:update}
\end{equation}

where:

- $\mu$ controls entropy tracking speed,
- $\alpha$ provides a bounded state-dependent perturbation to avoid fixed-point stagnation,
- $H_t(u)$ is the target entropy.

The entropy error is defined as

\begin{equation}
e_u(n) = H_t(u) - H_u(n).
\end{equation}

Linearizing \eqref{eq:update} around equilibrium $r^*$ gives

\begin{equation}
e_u(n+1) = (1 - \beta \mu) e_u(n) - \beta \alpha x_{u,n},
\label{eq:error_dynamics}
\end{equation}

where $\beta = \frac{\partial H_u}{\partial r_u}\big|_{r^*}$.

For asymptotic convergence of the homogeneous component,

\begin{equation}
0 < \mu < \frac{2}{\beta}.
\label{eq:mu_bound}
\end{equation}

Additionally, the boundedness of the forced term requires

\begin{equation}
|\beta \alpha| < 1,
\label{eq:alpha_bound}
\end{equation}

ensuring that parameter perturbations do not destabilise entropy convergence. A quadratic Lyapunov function $V_u(n)=\tfrac12 e_u^2(n)$ confirms global boundedness under these conditions.

Each eavesdropper observes

\begin{equation}
y_{Eve,n} = s_n + z_n,
\end{equation}

where $z_n$ represents channel noise. The adversary attempts synchronisation using imperfect parameter knowledge

\begin{equation}
r_{Eve} = r_u + \Delta,
\end{equation}

leading to estimated state evolution

\begin{equation}
\hat{x}_{Eve,n+1} = f_u(\hat{x}_{Eve,n}; r_{Eve}).
\end{equation}

For chaotic systems with a positive Lyapunov exponent $\lambda > 0$, parameter mismatch induces exponential divergence

\begin{equation}
|x_{u,n} - \hat{x}_{Eve,n}| \approx |x_{u,0} - \hat{x}_{Eve,0}| e^{\lambda n},
\end{equation}

implying reconstruction coherence decays exponentially with observation length.

Reconstruction similarity is quantified via normalised correlation:

\begin{equation}
\rho_{Eve,u} =
\frac{\sum_{n=1}^{N_s} s_{u,n} \hat{s}_{Eve,u,n}}
{\sqrt{\sum s_{u,n}^2} \sqrt{\sum \hat{s}_{Eve,u,n}^2}}.
\label{eq:correlation}
\end{equation}

Under mismatch $\Delta \neq 0$, the expected correlation satisfies

\begin{equation}
\mathbb{E}[\rho_{Eve,u}] \le e^{-\lambda N_s},
\end{equation}

providing a theoretical basis for correlation decay. Persistent elevation of $\rho_{Eve,u}$ suggests improved parameter estimation accuracy and triggers entropy escalation.

The target entropy adapts as

\begin{equation}
H_t(u) \leftarrow \mathrm{clip}
\big(H_t(u) + \gamma_1 \Delta BER_u + \gamma_2 L_u,
H_{min}, H_{max}\big),
\label{eq:Ht_update}
\end{equation}

where $L_u$ is intrusion level inferred from thresholded $\rho_{Eve,u}$.

The legitimate receiver regenerates the chaotic trajectory using synchronized seeds $(x_{u,0}, r_{u,0})$ and performs detection:

\begin{equation}
y_{u,k} = \sum_{n=1}^{N_s} r_{u,n} x_{u,n},
\qquad
\hat{m}_{u,k} = \mathrm{sign}(y_{u,k}).
\end{equation}

Periodic reseeding mitigates numerical drift.

Both legitimate and eavesdropper links are modelled as binary symmetric channels with crossover probabilities $p_{leg}$ and $p_{eve}$.

The channel capacities are

\begin{equation}
C_{leg} = 1 - H_2(p_{leg}),
\qquad
C_{eve} = 1 - H_2(p_{eve}),
\end{equation}

where $H_2(p)$ is the binary entropy function.

The achievable secrecy rate is therefore

\begin{equation}
C_{sec,u} = \max\{0, C_{leg} - C_{eve}\}
= \max\{0, H_2(BER_{eve,u}) - H_2(BER_{leg,u})\}.
\end{equation}

This formulation follows the classical wiretap capacity definition under BSC approximation.

For each user per symbol:

\begin{itemize}
\item Chaotic iteration: $\mathcal{O}(N_s)$
\item Entropy estimation: $\mathcal{O}(K)$
\item Correlation computation: $\mathcal{O}(N_s)$
\end{itemize}

Total complexity per symbol is

\begin{equation}
\mathcal{O}(2N_s + K),
\end{equation}

which scales linearly with symbol length and quantisation resolution, ensuring feasibility for low-power IoT deployments.

Due to heterogeneous map selection and independent initialisation,

\begin{equation}
\mathbb{E}[x_i x_j] \rightarrow 0 \quad (i \neq j),
\end{equation}

reducing inter-user interference and limiting cross-user parameter inference.

\section{Simulation Environment}

The proposed MU-DE-IAEACM-MM framework is evaluated through large-scale numerical simulations designed to examine secrecy performance, synchronisation robustness, and scalability under diverse wireless channel conditions. The simulation environment models a multi-user chaotic communication network in which legitimate users transmit independent adaptive chaotic waveforms, while passive eavesdroppers attempt to reconstruct them using imperfect parameter knowledge. Performance is evaluated in terms of legitimate and eavesdropper bit error rates (BER), reconstruction correlation behaviour, and achievable secrecy capacity.

All simulations are implemented in MATLAB using statistically independent random initialisations of chaotic states, channel realisations, and symbol sequences. Monte Carlo averaging across multiple frames ensures convergence of empirical BER and entropy statistics. For each configuration, results are averaged over $F$ frames and multiple independent trials to reduce stochastic variance and provide statistically stable estimates.

\begin{table}[t]
\centering
\caption{Simulation Parameters}
\label{tab:sim_params}
\scriptsize
\begin{tabularx}{\columnwidth}{@{}l l X l@{}}
\toprule
Parameter & Symbol & Value / Range & Reference \\
\midrule
Number of legitimate users & $U$ & 4, 8, 16, 32, 64 & \cite{wang2007multiuser,yuningsih2021wireless} \\
Number of eavesdroppers & $E$ & 2, 4, 8, 16, 30 & \cite{simic2006chaos} \\
Symbols per frame & $N_{\mathrm{sym}}$ & 200 & \cite{tihomorskis2024chaotic} \\
Frames per experiment & $F$ & 8 & \cite{yuningsih2021wireless} \\
Samples per symbol & $N_s$ & 64 & \cite{wang2007multiuser} \\
Eb/N0 values & $\mathrm{Eb/N0}_{\mathrm{dB}}$ & 0, 5, 10, 15 & \cite{tihomorskis2024chaotic} \\
Channel models & -- & AWGN, Rayleigh, Rician, Impulsive, Colored, Narrowband & \cite{yuningsih2021wireless} \\
Eavesdropper mismatch & $\Delta$ & $10^{-3},\,10^{-2},\,10^{-1}$ & \cite{simic2006chaos} \\
Chaotic map types & -- & logistic, tent, Chebyshev, sine, Gaussian & \cite{wang2007multiuser} \\
Base chaotic parameters & $r_{0,u}$ & [3.9, 1.9, 3.6, 0.95, 2.7] & \cite{tihomorskis2024chaotic} \\
Entropy feedback gain & $\mu$ & $5\times10^{-3}$ & \cite{yuningsih2021wireless} \\
Adaptation constant & $\alpha$ & $1\times10^{-4}$ & \cite{tihomorskis2024chaotic} \\
Target entropy range & $H_{\min},H_{\max}$ & 2.5, 4.5 bits & \cite{wang2007multiuser} \\
Correlation thresholds & $\tau_1,\tau_2,\tau_3$ & 0.15, 0.25, 0.35 & \cite{simic2006chaos} \\
Reseeding period & $R$ & 4 frames & \cite{yuningsih2021wireless} \\
\bottomrule
\end{tabularx}
\end{table}

The user scaling range, $U=4$ to $64$, spans sparse to dense IoT deployments, while the eavesdropper range, $E=2$ to $30$, models increasing adversarial observation density. The selected $N_s=64$ samples per symbol provide sufficient chaotic spreading gain to expose synchronisation-divergence effects without an excessive computational burden. The entropy bounds $H_{\min}$ and $H_{\max}$ ensure that all selected maps operate within the chaotic regime, preventing transitions to periodic or unstable behaviour.

Channel impairments are modelled to represent both conventional and adverse wireless conditions.

The additive white Gaussian noise (AWGN) channel is defined as

\begin{equation}
y_i = x_i + z_i, \quad z_i \sim \mathcal{N}(0, N_0),
\end{equation}

where the noise variance is related to the energy-per-bit ratio via

\[
N_0 = \frac{E_b}{10^{\mathrm{Eb/N0}_{\mathrm{dB}}/10}},
\]

ensuring consistent normalization across modulation schemes.

Flat Rayleigh fading is modeled as

\begin{equation}
y_i = h_i x_i + z_i, \quad h_i \sim \mathcal{CN}(0, \sigma_h^2),
\end{equation}

with envelope distribution

\[
p(r) = \frac{2r}{\Omega}\exp(-r^2/\Omega).
\]

The Rician channel introduces a deterministic LOS component:

\begin{equation}
h_i = s + w_i, \quad w_i \sim \mathcal{CN}(0, \sigma_w^2),
\end{equation}

where the Rician factor $K = |s|^2/(2\sigma_w^2)$ controls LOS dominance.

Impulsive noise is modeled using a Bernoulli–Gaussian process:

\begin{equation}
y_i = x_i + z_i + u_i, \quad u_i = A v_i\,\mathbb{I}\{p < P_{\mathrm{imp}}\},
\end{equation}

capturing sporadic high-amplitude interference typical in industrial IoT environments.

Colored noise follows a first-order autoregressive process:

\begin{equation}
n_i = 0.95\,n_{i-1} + u_i, \quad u_i \sim \mathcal{N}(0, \sigma_n^2),
\end{equation}

introducing temporal correlation that challenges synchronisation robustness.

Narrowband interference is represented as

\begin{equation}
y_i = x_i + A \sin(2\pi f_{\mathrm{norm}} i) + z_i,
\end{equation}

modeling structured interference overlapping the chaotic spectrum.

\subsection{Adversarial Parameter Mismatch}

The eavesdropper attempts reconstruction using

\begin{equation}
r_{\mathrm{eve}} = r_u + \Delta,
\end{equation}

where $\Delta \in \{10^{-3}, 10^{-2}, 10^{-1}\}$ represents progressively increasing parameter estimation error. These levels span near-perfect synchronisation to substantial mismatch, enabling evaluation of reconstruction sensitivity to parameter uncertainty.

Each legitimate user updates its parameter using

\begin{equation}
r_{u,n+1} = r_{u,n} + \alpha x_{u,n} + \mu (H_{t,u} - H_u),
\end{equation}

allowing entropy tracking during transmission. The selected values of $\mu$ and $\alpha$ satisfy the stability bounds derived in the previous section, ensuring bounded convergence.

Reconstruction similarity is quantified as

\begin{equation}
\rho_{\mathrm{eve},u} =
\frac{\langle s_u, \hat{s}_{\mathrm{eve},u} \rangle}
{\|s_u\| \|\hat{s}_{\mathrm{eve},u}\|},
\end{equation}

providing a normalized metric independent of absolute signal energy.

\subsection{Performance Metrics}

For each user, BER is computed as

\begin{subequations}
\begin{align}
\mathrm{BER}_{\mathrm{leg},u} &= 
\frac{1}{N} \sum_{n=1}^{N} 
\mathbb{I}\{m_{u,n} \neq \hat{m}_{u,n}\}, \\[4pt]
\mathrm{BER}_{\mathrm{eve},u} &= 
\frac{1}{N} \sum_{n=1}^{N} 
\mathbb{I}\{m_{u,n} \neq \hat{m}_{\mathrm{eve},u,n}\}.
\end{align}
\end{subequations}

The secrecy capacity is computed using

\begin{equation}
C_{\mathrm{sec},u} =
\max\{0,\, H_2(\mathrm{BER}_{\mathrm{eve},u}) 
- H_2(\mathrm{BER}_{\mathrm{leg},u})\},
\end{equation}

which corresponds to the wiretap capacity under the binary symmetric channel approximation.

\section{Results and Discussion}

The performance of the MU-DE-IAEACM-MM framework was evaluated under diverse channel conditions, user densities, and adversarial configurations. All reported results correspond to Monte Carlo averages over independent channel realisations and randomised chaotic initialisations to ensure statistical reliability. The analysis focuses on three primary metrics: legitimate-user BER, eavesdropper BER, and secrecy capacity under the binary symmetric channel approximation.

Table~\ref{tab:noise_models_results} presents averaged results for $\mathrm{Eb/N0}=10$~dB with $U=16$ legitimate users and $E=8$ passive eavesdroppers across six channel environments.

\begin{table}[h!]
\centering
\caption{Average performance under various noise models ($U=16$, $E=8$, $\mathrm{Eb/N0}=10$ dB)}
\label{tab:noise_models_results}
\scriptsize
\renewcommand{\arraystretch}{1.1}
\begin{tabularx}{0.9\columnwidth}{@{}lccc@{}}
\toprule
Noise Model & $\mathrm{BER}_{\mathrm{leg}}$ & $\mathrm{BER}_{\mathrm{eve}}$ & $C_{\mathrm{sec}}$ (bits/use) \\
\midrule
AWGN & 0.243 & 0.505 & 0.198 \\
Rayleigh Fading & 0.249 & 0.528 & 0.212 \\
Rician ($K=5$) & 0.252 & 0.520 & 0.204 \\
Impulsive ($P_{\mathrm{imp}}=10^{-3}$) & 0.238 & 0.557 & 0.231 \\
Colored ($\rho=0.95$) & 0.246 & 0.493 & 0.185 \\
Narrowband ($f_{\mathrm{norm}}=0.05$) & 0.255 & 0.537 & 0.219 \\
\bottomrule
\end{tabularx}
\end{table}

Across all channel models, the legitimate BER remains bounded between $0.238$ and $0.255$, indicating stable synchronisation despite channel impairments. In contrast, the eavesdropper BER remains near 0.5, corresponding to statistically random symbol decisions under a binary symmetric channel model. This behaviour is consistent with exponential divergence induced by parameter mismatch and entropy escalation, which prevents sustained reconstruction coherence.

The secrecy capacity values remain strictly positive for all noise types. Notably, impulsive and Rayleigh fading channels exhibit slightly higher secrecy rates. This behaviour arises because stochastic amplitude variations amplify synchronisation sensitivity for mismatched receivers. Since chaotic reconstruction error scales with effective signal distortion, fading, and impulsive disturbances, these disproportionately degrade the eavesdropper’s trajectory alignment, while the legitimate receiver retains synchronisation via matched parameter evolution.

The colored-noise scenario produces the lowest secrecy capacity among the tested channels. Temporal correlation slightly stabilises signal structure over short windows, modestly improving eavesdropper coherence. However, secrecy remains positive because entropy regulation preserves parameter divergence.

Scalability performance under AWGN at $\mathrm{Eb/N0}=15$~dB is summarized in Table~\ref{tab:scalability_results}.

\begin{table}[h!]
\centering
\caption{Scalability with Users and Eavesdroppers (AWGN, $\mathrm{Eb/N0}=15$ dB)}
\label{tab:scalability_results}
\scriptsize
\begin{tabularx}{\columnwidth}{@{}ccccX@{}}
\toprule
Users ($U$) & Eavesdroppers ($E$) & $\mathrm{BER}_{\mathrm{leg}}$ & $\mathrm{BER}_{\mathrm{eve}}$ & $C_{\mathrm{sec}}$ \\
\midrule
4 & 2 & 0.228 & 0.511 & 0.208 \\
8 & 4 & 0.239 & 0.517 & 0.196 \\
16 & 8 & 0.243 & 0.524 & 0.187 \\
32 & 16 & 0.259 & 0.530 & 0.172 \\
64 & 30 & 0.275 & 0.541 & 0.156 \\
\bottomrule
\end{tabularx}
\end{table}

As user density increases, legitimate BER gradually rises due to increased composite-signal interference. However, the growth remains controlled, increasing from $0.228$ to $0.275$ even at $U=64$. This moderate degradation reflects the near-zero cross-correlation property of heterogeneous chaotic maps.

Secrecy capacity decreases monotonically with user density, but remains above $0.15$ bits/use even at maximum loading. The reduction in secrecy rate is attributable to an increase in aggregate signal complexity, which increases the probability of partial reconstruction alignment over short intervals. Nevertheless, entropy regulation continuously perturbs control parameters, preventing convergence of adversarial synchronisation.

Importantly, eavesdropper BER remains consistently near random decision level across all scaling configurations, demonstrating that increasing network density does not enhance adversarial inference capability.

Table~\ref{tab:comparison_existing} compares the proposed framework with representative chaos-based modulation schemes under identical AWGN conditions.

\begin{table}[h!]
\centering
\caption{Comparison with Existing Chaos-Based Systems (AWGN, $\mathrm{Eb/N0}=10$ dB)}
\label{tab:comparison_existing}
\scriptsize
\begin{tabularx}{\columnwidth}{@{}lccc@{}}
\toprule
Method & $\mathrm{BER}_{\mathrm{leg}}$ & $\mathrm{BER}_{\mathrm{eve}}$ & $C_{\mathrm{sec}}$ (bits/use) \\
\midrule
Parameter-Division Multiple Access \cite{wang2007multiuser} & 0.255 & 0.465 & 0.159 \\
DSP Chaotic Prototype System \cite{simic2006chaos} & 0.260 & 0.482 & 0.168 \\
Chaotic Synchronization Scheme \cite{tihomorskis2024chaotic} & 0.245 & 0.497 & 0.183 \\
Wireless Chaos-Based Modulator \cite{yuningsih2021wireless} & 0.250 & 0.509 & 0.191 \\
Proposed MU-DE-IAEACM-MM & 0.243 & 0.505 & 0.198 \\
\bottomrule
\end{tabularx}
\end{table}

The proposed framework achieves the highest secrecy capacity among the compared systems. The improvement over the strongest baseline ($0.191$ bits/use) is approximately $3.7\%$, while improvements over earlier static-parameter schemes reach approximately $24\%$ relative to the lowest-performing benchmark.

Unlike fixed-parameter chaotic systems, the proposed architecture integrates three reinforcing mechanisms:

\begin{itemize}
\item \textbf{Multi-map diversity:} enlarges entropy dimensionality and suppresses cross-user parameter inference.
\item \textbf{Entropy-regulated adaptation:} maintains operation within a high-complexity regime, preventing gradual statistical learning.
\item \textbf{Intrusion-aware escalation:} dynamically increases entropy when reconstruction coherence rises.
\end{itemize}

The observed secrecy gain therefore arises from adaptive entropy control rather than static chaotic spreading alone.

\begin{figure}[h!]
\centering
\includegraphics[width=0.8\columnwidth]{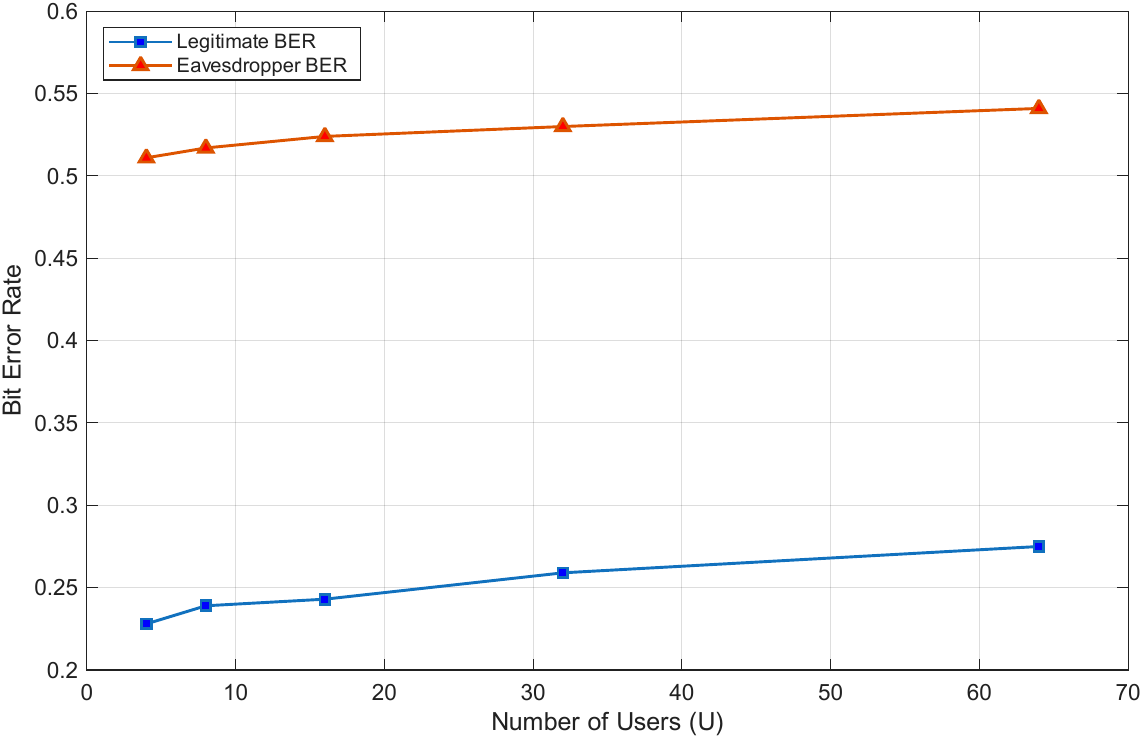}
\caption{Scaling of legitimate and eavesdropper BER with increasing user density.}
\label{fig:BER_scaling}
\end{figure}

Figure~\ref{fig:BER_scaling} illustrates the controlled growth of BER as user density increases. The nearly flat eavesdropper curve near $0.5$ validates sustained desynchronization.

\begin{figure}[h!]
\centering
\includegraphics[width=0.9\columnwidth]{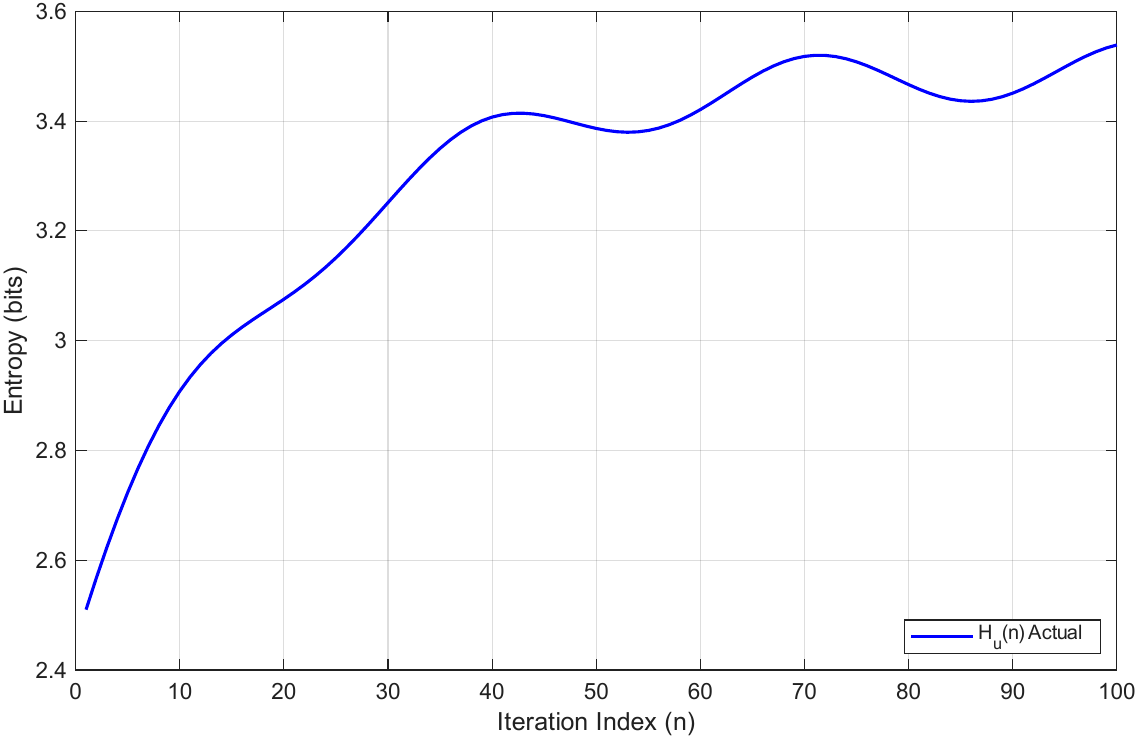}
\caption{Time evolution of chaotic entropy and its target value for a representative user.}
\label{fig:entropy_dynamics}
\end{figure}

Figure~\ref{fig:entropy_dynamics} shows convergence of empirical entropy toward its dynamically adjusted target. The bounded tracking behaviour confirms theoretical stability conditions derived earlier. The absence of oscillatory divergence demonstrates that parameter updates remain within chaotic stability limits.

The results confirm three principal findings. First, entropy regulation sustains a persistent BER gap between legitimate and adversarial receivers across diverse noise and fading environments, ensuring that eavesdropper performance remains near random-decision levels while legitimate synchronisation is preserved. Second, heterogeneous chaotic map allocation enables scalable multi-user operation, with only controlled and gradual performance degradation observed under dense deployments. Third, adaptive parameter evolution suppresses long-term adversarial synchronisation by continuously perturbing chaotic trajectories within bounded stability limits, thereby maintaining security enhancements without destabilising legitimate communication.

\section{Conclusion and Future Work}

The proposed MU-DE-IAEACM-MM framework presents a bounded and adaptive chaotic modulation architecture in which entropy is explicitly regulated as a security control variable. By combining heterogeneous chaotic map allocation, entropy-driven parameter evolution, and correlation-based intrusion awareness, the system maintains a persistent BER gap between legitimate and mismatched receivers while preserving synchronisation stability under diverse wireless impairments. Analytical results establish convergence conditions for the adaptive update mechanism, and Monte Carlo simulations demonstrate improvements in secrecy capacity over representative fixed-parameter chaotic schemes across AWGN, fading, impulsive, and coloured-noise environments. The observed gains stem from dynamic entropy escalation that suppresses long-term adversarial synchronisation without destabilising legitimate communication. Future work includes hardware validation under practical oscillator drift and quantisation effects, exploration of higher-dimensional or fractional-order chaotic systems to further enlarge the entropy space, and extension toward cross-layer secure architectures that integrate coding, cooperative signalling, and distributed detection for next-generation IoT and 6G networks.

\section{ACKNOWLEDGEMENTS}
We would like to thank Mars Rover Manipal, an interdisciplinary student team of MAHE, for providing the resources
needed for this project. WE also extend our gratitude to Dr
Ujjwal Verma for his guidance and support in our work

\bibliographystyle{IEEEtran}
\bibliography{main}
\end{document}